\documentclass[showpacs,aps,amsmath,amssymb,twocolumn]{revtex4}
\usepackage[latin1]{inputenc}
\usepackage{dcolumn}
\usepackage{bm}
\usepackage{verbatim} 
\usepackage[]{graphicx}			
\usepackage{color}			
\usepackage{subfig}			
\graphicspath{{./figuras/}}		
\usepackage{caption}	
\newcommand{\be}{\begin{equation}}
\newcommand{\ee}{\end{equation}}
\newcommand{\ben}{\begin{eqnarray}}
\newcommand{\een}{\end{eqnarray}}
\newcommand{\bes}{\begin{subequations}}
\newcommand{\ees}{\end{subequations}}
\newcommand{\bb}{\bibitem}

\newcommand{\nn}{\nonumber\\}
\newcommand{\bfi}{\begin{figure}}
\newcommand{\efi}{\end{figure}}
\newcommand{\bc}{\begin{center}}
\newcommand{\ec}{\end{center}}

\begin{document}
\title{First-order formalism for twinlike models with several real scalar fields}

\author{D. Bazeia$^{1,2}$, A.S. Lob\~ao Jr.$^{1}$, L. Losano$^{1,2}$ and R. Menezes$^{2,3}$}

\affiliation{$^1$Departamento de F\'\i sica, Universidade Federal da Para\'\i ba, 58051-970 Jo\~ao Pessoa, PB, Brazil\\ $^2$Departamento de F\'\i sica, Universidade Federal de Campina Grande, 58109-970 Campina Grande, PB, Brazil and \\ $^3$Departamento de Ci\^encias Exatas, Universidade Federal da Para\'\i ba, 58297-000 Rio Tinto, PB, Brazil.}

\date{\today}

\begin{abstract}
We investigate the presence of twinlike models in theories described by several real scalar fields. We focus on the first-order formalism, and we show how to build distinct scalar field theories that support the same extended solution, with the same energy density and the very same linear stability. The results are valid for two distinct classes of generalized models, that include the standard model and cover a diversity of generalized models of current interest in high energy physics. 
\end{abstract}


\maketitle

\section{introduction}

Kinks, vortices and monopoles are defect structures that play interesting role in high energy physics and have been studied in a diversity of scenarios \cite{a1,a2}. Vortices and monopoles in general require the presence of gauge fields, Abelian and non Abelian, respectively. However, in the case
of models described by a real scalar field $\phi$ in two  spacetime dimensions, with $x^\mu=(x^0=t,x^1=x)$ and $x_\mu=(x_0=t,x_1=-x)$, the defect structures represent static configurations $\phi=\phi(x)$ known as kinks, describing solutions of the equation of motion with the  asymptotic profile $\phi(x\to\infty)\neq\phi(x\to-\infty)$. 

In models with standard kinematics, the kink profile is controlled by the potential $V=V(\phi)$, which usually engenders spontaneous symmetry breaking.
However, kinklike structures may also appear in generalized models, where the kinematics is modified from the standard one, allowing for the derivative of the field to appear in a generalized way, which we further explain below. One sometimes refers to such generalized models as k-field models \cite{K}, which were introduced with the main motivation to help us to understand the current accelerated expansion of the Universe.

The generalized models open new routes and introduce a diversity of issues, among them the interesting possibility that two distinct models, one standard and the other generalized, could support the same kinklike solution, with the very same energy density \cite{tro}. These models are called twinlike models, and several investigations on the issue have been introduced recently \cite{PBSC}. In these investigations, one could identify interesting twinlike models, having the same kinklike solution, with the same energy density and the very same linear stability.

In the current work we deal with kinks in models described by several real scalar fields, and we focus on the twinlike issue, that is, on the presence of distinct models describing the very same kinklike solution, with the same energy density and possibly the same linear stability. We concentrate mainly on the formal aspects one needs to obtain twinlike models with standard and generalized kinematics, and we illustrate the results with examples of current interest in high energy physics. Due to the complexity of the subject, we search for kinklike structures and study the corresponding linear stability, using the first-order formalism, with very much help us to reach the general results of the current work. We stress here that the first-order formalism refers to first-order differential equations, whose solutions solve the equations of motion; it is a procedure to find exact solutions, and it has nothing to do with any pertutbative procedure. For this reason, in the next Section we start presenting the first-order formalism for a generic model, containing several real scalar fields, with generalized kinematics. This investigation reviews and generalizes previous work on the subject \cite{BM}. It also shows that it is not a simple task to go explicitly to the first-order framework and find analytical solutions \cite{X}, an issue related to supersymmetry, to be considered elsewhere under the general guidance of Ref.~\cite{Y}. The next step is then to deal with twinlike models, and this is done in Sec.~III. There we introduce two distinct routes to study the subject, including the corresponding linear stability.  We end the work in Sec.~IV, where we present our comments and conclusions.

The current study concerns the presence of defect structures in generalized models with several real scalar fields, so it of direct interest to cosmology, to provide alternative descriptions of k-field theories \cite{K,tro} in the presence of several fields, a subject of direct interest to multifield inflation and multifield defect networks, as one finds, for instance, in Ref.~\cite{X,Z}. The models that we investigate also engender generic properties of string theory, and as such they provide another well-motivated subject of interest in high energy physics.

\section{First-order formalism}

In this Section we focus on issues that review and generalize the first-order formalism previously introduced in Ref.~\cite{BM}. We deal with several scalar fields with generalized kinematics, and the models that we investigate are described by the generic action, containing $N$ real scalar fields $\{\phi_i;\, i=1,2,\ldots,N\}$ in the two-dimensional
space-time:
\be
{\cal S}=\int d^2 x \; {\cal L}(\phi_i,X_{ij})\,,
\ee
where 
\be\label{x}
X_{ij}=\frac12\partial_{\mu}\phi_i \partial^{\mu} \phi_j \,.
\ee
We use dimensionless units, where the scalar fields, space and time coordinates, and coupling constants are all dimensionless. 

The energy-momentum tensor has the form
\be
T_{\mu\nu}={\cal L}_{X_{ij}}  \partial_{\mu} \phi_i \partial_{\nu} \phi_j - g_{\mu\nu} {\cal L}\,,
\ee
where we are using the standard notation: ${\cal L}_{A}=\partial{\cal L}/\partial A$, etc.

There are $N$ equations of motion; they are given by
\be\label{eqm1}
 \partial_{\mu} \left({\cal L}_{X_{ij}}  \partial^{\mu} \phi_j\right)={\cal L}_{\phi_i}\,.
\ee
We can rewrite them as
\be
{\cal G}_{ij}^{\alpha\beta}\partial_{\alpha} \partial_{\beta} \phi_j + 2 X_{jl}{\cal L}_{X_{ij}\phi_l} - {\cal L}_{\phi_i}=0\,,
\ee
where 
\be
{\cal G}^{\alpha\beta}_{ij}={\cal L}_{X_{ij}} g^{\alpha\beta} + {\cal L}_{X_{il}X_{jm}} \partial^{\alpha} \phi_l \partial^{\beta} \phi_m\,.
\ee

We search for defect structures, so we consider the case of static fields. We write $\phi_i=\phi_i(x)$, and the $N$ equations of motion now become
\be\label{eqm2}
({\cal L}_{X_{ij}}+2 {\cal L}_{X_{il}X_{jm}} X_{lm}) \phi_j^{\prime\prime} -2 X_{jl} {\cal L}_{X_{ij}\phi_l} +{\cal L}_{\phi_i}=0\,,
\ee
where the prime denotes derivative with respect to $x$ and $X_{ij}=-\phi_i^\prime\phi_j^\prime/2$.  
These equations can be integrated once to give
\be
{\cal L}-2 {\cal L}_{X_{ij}}X_{ij}=0.
\ee
In the above equation we have discarded an integration constant, in order to ensure stability of the defect structures; as one knows, the vanishing of the integration constant corresponds to making the static solutions stressless, obeying: $\tau(x)=T_{11}=0$.

The energy density of the static solutions can be written as
\be
\rho(x)=T_{00} =-{\cal L}= {\cal L}_{X_{ij}}\phi_i^\prime\phi_j^\prime\,.
\ee 
At this stage, we introduce a new function of the several fields; we call it $W=W(\phi_1,\phi_2,\ldots,\phi_n)$ and we write
\be\label{Lw}
{\cal L}_{X_{ij}}\phi_j^\prime=W_{\phi_i}\,.
\ee
This is important because it allows obtaining the energy density as a total derivative, $\rho(x)=dW/dx$, such that the energy can be written as
\ben
E=\Delta W&=&|W\left(\phi_1(\infty),\phi_2(\infty),\ldots,\phi_n(\infty)\right)\nonumber\\
&-&W\left(\phi_1(-\infty),\phi_2(-\infty),\ldots,\phi_n(-\infty)\right)|\,.\nonumber\\ 
&&
\een
Moreover, we substitute \eqref{Lw} in the equations \eqref{eqm2} to get the set of $N$ first-order differential equations 
\be
W_{\phi_i,\phi_j}\phi_j^{\prime}=-{\cal L}_{\phi_i}\,,
\ee
which also solve the equations of motion. Thus, the choice \eqref{Lw} leads us to the first-order formalism, that is, we now solve the equations of motion solving the first-order differential equations above.

Let us now examine linear stability.  We consider $\phi_i(x,t)=\phi_i(x)+\eta_i(x,t)$, where $\eta_i(x,t)$ are small fluctuations around the static solution. In this case, up to first-order in the fluctuations  we have
\be
X_{ij}\rightarrow X_{ij}+\bar{X}_{ij}\,,
\ee
where
\be
\bar{X}_{ij}=\frac12\partial_{\mu} \phi_i \partial^{\mu} \eta_j+\frac12\partial_{\mu} \phi_j\partial^{\mu} \eta_i\,,
\ee
such that
\bes
\be
{\cal L}_{\phi_i}\rightarrow{\cal L}_{\phi_i}+{\cal L}_{\phi_i\phi_j}\eta_j+{\cal L}_{\phi_i X_{jk}}\bar{X}_{jk}\,,
\ee
and
\be
{\cal L}_{X_{ij}}\rightarrow{\cal L}_{X_{ij}}+{\cal L}_{X_{ij}\phi_k}\eta_k+{\cal L}_{X_{ij} X_{ml}}\bar{X}_{ml}\,.
\ee
\ees
We put these expressions into the equation of motion \eqref{eqm1} to obtain
\ben\label{st01}
&& \left({\cal L}_{X_{mi}X_{lj}}\partial_{\mu} \phi_m\partial_{\alpha} \phi_l+ {\cal L}_{X_{ij}}g_{\mu\alpha}\right)\partial^{\mu}\partial^{\alpha}\eta_i \nonumber\\
&&+\big[\partial^{\mu} \left({\cal L}_{X_{mi}X_{lj}}\partial_{\mu} \phi_m\partial_{\alpha} \phi_l+ {\cal L}_{X_{ij}}g_{\mu\alpha}\right)\nonumber\\
&&-\left({\cal L}_{X_{il}\phi_{j}}-{\cal L}_{X_{lj}\phi_{i}}\right)g_{\mu\alpha}\partial^{\mu}\phi_l\big]\partial^{\alpha}\eta_i\nonumber\\
&&+\big[\partial^{\mu} \left({\cal L}_{X_{lj}\phi_{i}}\partial_{\mu} \phi_l\right)-{\cal L}_{\phi_{i}\phi_{j}}\big]\eta_i=0\,,
\een
which for static solutions reduces to
\ben\label{st02}
&&{\cal L}_{X_{ij}} \square\eta_i-2X_{ml} {\cal L}_{X_{mi}X_{lj}}\eta_i^{\prime\prime}\nonumber\\
&&-\left[\left({\cal L}_{X_{ij}}+2X_{ml}{\cal L}_{X_{mi}X_{lj}}\right)^\prime-\left({\cal L}_{X_{il}\phi_{j}}-{\cal L}_{X_{lj}\phi_{i}}\right)\phi_l^\prime\right]\eta_i^\prime\nonumber\\
&&-\big[\left({\cal L}_{X_{lj}\phi_{i}}\phi_l^\prime\right)^\prime+{\cal L}_{\phi_{i}\phi_{j}}\big]\eta_i=0\,,\nn
&&
\een
where $\square$ is the D'Alambertian operator.

Now, from \eqref{st02} and using
\be
\eta_i(x,t)=\eta_i(x)\cos(\omega t)\,,
\ee
we have
\ben\label{stab1}
&&-\left[\left({\cal L}_{X_{ij}}+2X_{ml}{\cal L}_{X_{mi}X_{lj}} \right)\eta_i^{\prime}\right]^{\prime}\nonumber\\
&&+\left({\cal L}_{X_{il}\phi_{j}}-{\cal L}_{X_{lj}\phi_{i}}\right)\phi_l^\prime\,\eta_i^\prime\nonumber\\
&&=\big[\omega^2 {\cal L}_{X_{ij}}+{\cal L}_{\phi_{i}\phi_{j}}+\left({\cal L}_{X_{lj}\phi_{i}}\phi_l^\prime\right)^\prime\big]\eta_i\,,
\een
which has the general form
\begin{equation}\label{sch_mod}
-a_{ij}\eta_i^{\prime\prime}-\left(a_{ij}^{\prime}+b_{ij}\right)\eta_i^{\prime}-c_{ij}\eta_i=\omega^2{\cal L}_{X_{ij}}\eta_i\,,
\end{equation}
where
\begin{eqnarray*}
a_{ij}&=&{\cal L}_{X_{ij}}+2X_{ml}{\cal L}_{X_{mi}X_{lj}}\,,\\
b_{ij}&=&-\left({\cal L}_{X_{il}\phi_{j}}-{\cal L}_{X_{lj}\phi_{i}}\right)\phi_l^\prime\,,\\
c_{ij}&=&{\cal L}_{\phi_i\phi_j}+({\cal L}_{X_{lj}\phi_{i}}\phi_l^\prime)^{\prime}\,.
\end{eqnarray*}

We can modify the equation \eqref{sch_mod} into the Schroedinger-like equation
\begin{equation}
\left(-\delta_{ij}\frac{d^2}{dz^2}+U_{ij}\right)u_i=\omega^2 u_j\,,
\end{equation}
where the potential $U$ is now a matrix which depends on the matrix $S$ and $R$, introduced as folows: in \eqref{sch_mod} we change $\eta_i (x)$ by $u_i(z)$, such that
\begin{eqnarray}
\eta_j=S_{jk}u_k \mbox{\;\;\;\;\; and \;\;\;\;\;}  dx=\frac{dz}{R}\,.
\end{eqnarray}
In this case, the Schroedinger-like equation requires that
\begin{subequations}
\begin{eqnarray}
2a_{ij}R\frac{dS_{jk}}{dz}+\frac{d(a_{ij}R)}{dz}S_{jk}+b_{ij}S_{jk}&=&0\,,\\
R^{-2}S_{lm}^{-1}a_{li}^{-1}{\cal L}_{X_{ij}}S_{jk}&=&\delta_{mk}\,.
\end{eqnarray}
\end{subequations}

Let us now examine a simpler but important situation. To implement this, we consider the general model described by 
\be\label{L2}
{\cal L}={\cal L}({\cal X},\phi_i)\,,
\ee
where $i,j=1,2,\ldots,N$, and ${\cal X}=\delta_{ij}{X_{ij}}$,
with ${X_{ij}}$ given by  \eqref{x}. In the case of static solutions $\phi_i=\phi_i(x)$, the equations of motion and the stressless condition are given by
\be\label{eqm1i}
-[{\cal L}_{\cal X}\phi_i^\prime]^\prime={\cal L}_{\phi_i}\,,
\ee
and
\be
{\cal L}-2{\cal L}_{\cal X}{\cal X}=0\,.
\ee
The $N$ first-order equations \eqref{Lw} have the form
\be\label{Lw2}
{\cal L}_{\cal X}\phi_i^\prime=W_{\phi_i}\,.
\ee
They lead us to
\be
\frac{d\phi_i}{d\phi_j}=\frac{W_{\phi_i}}{W_{\phi_j}}\,,
\ee
which can be used to find explicit solutions, for Lagrange densities of the form \eqref{L2}.

In general, the energy-momentum tensor is given by
\begin{equation}
T_{\mu\nu}=-g_{\mu\nu}{\cal L}+{\cal L}_{\cal X}\partial_\mu {\phi_i}\partial_\nu{\phi_i}\,,
\end{equation}
and for static solutions we have
\bes\label{en_tensor}
\begin{eqnarray}
T_{00}&=&-{\cal L} ,\\
T_{11}&=&{\cal L}+{\cal L}_{\cal X}\phi_i^{\prime}\phi_i^{\prime}\,.
\end{eqnarray}
\ees
Also, the stability equation \eqref{stab1} can be written as
\begin{equation}\label{stab2}
-a_{ij}\eta_i^{\prime\prime}-\left(a_{ij}^{\prime}+b_{ij}\right)\eta_i^{\prime}-c_{ij}\eta_i=\omega^2{\cal L}_{X_{ij}}\eta_i\,,
\end{equation}
where, now
\begin{eqnarray*}
a_{ij}&=&\delta_{ij}{\cal L}_{\cal X}-\delta_{ki}{\cal L}_{\cal X X}\phi_k^{\prime}\phi_j^{\prime}\,,\\
b_{ij}&=&-\left(\delta_{ki}{\cal L}_{\phi_j{\cal X}}-\delta_{kj}{\cal L}_{\phi_i{\cal X}}\right)\phi_k^{\prime}\,,\\
c_{ij}&=&\left({\cal L}_{{\cal X}\phi_i}\phi_j^{\prime}\right)^{\prime}+{\cal L}_{\phi_i\phi_j}\,.
\end{eqnarray*}

In the case of  standard kinematics, the Lagrangian takes the form
\be\label{lsm}
{\cal L}={\cal X}-V(\phi_i)\,,
\ee
and the equations of motion \eqref{eqm1i} become
\be\label{eqm2i}
\phi_i^{\prime\prime}=V_{\phi_i}\,.
\ee
Moreover, the equations \eqref{en_tensor} give
\bes
\begin{eqnarray}
T_{00}&=&\frac{1}{2}\phi_i^{\prime}\phi_i^{\prime}+V({\phi_i}),\label{t00}\\
T_{11}&=&\frac{1}{2}\phi_i^{\prime}\phi_i^{\prime} -V({\phi_i})\label{t11}\,.
\end{eqnarray}
\ees
Moreover,  the first-order equations \eqref{Lw2} can be written as 
\be\label{eq4}
\phi_i^{\prime}=W_{\phi_i}\,,
\ee 
which, combined with the stressless conditions $T_{11}=0$, eq.\eqref{t11}, allows writing 
\be\label{vsm}
V=\frac12 W_{\phi_i}W_{\phi_i}\,.
\ee

We use this to rewrite the equations of motion $(\ref{eqm2i})$ as
\begin{equation}\label{eq3}
\phi_i^{\prime\prime}=W_{\phi_j}W_{\phi_j\phi_i}\,,
\end{equation}
This equation can be integrated once, and we obtain
\begin{equation}\label{t11c}
\phi_i^{\prime}\phi_i^{\prime} -W_{\phi_i}W_{\phi_i}=C\,,
\end{equation}
where $C$ is a constant that can be identified with the stress component, that is, $T_{11}=C$. Stability of  the static solution imposes that $C=0$, and the solutions are stressless. 
This changes the energy density $T_{00}$ to the form
\begin{equation}\label{eq5}
\rho(x)=\phi_i^{\prime}W_{\phi_i}=W_{\phi_i}W_{\phi_i}=\frac{dW}{dx}.
\end{equation}  
Thus, the energy associated with these configurations are given by
\begin{eqnarray}\label{eq6}
E&=&|W[\phi_1(\infty),\cdots,\phi_n(\infty)]\nonumber\\
&&-W[\phi_1(-\infty),\cdots,\phi_n(-\infty)]|\,.
\end{eqnarray}
In addition, the stability equations \eqref{stab2} became
\ben\label{stabi_stand}
-\eta_i^{\prime\prime}+V_{\phi_i\phi_j}\eta_j=\omega^2\eta_i\,.
\een

In order to illustrate the general investigation, let us now consider the specific model
\be
{\cal L}={\cal X}\left|{\cal X}\right|-V(\phi_i)\,.
\ee
The equations of motion are
\be
\phi_i^{\prime\prime}\phi_j^{\prime}\phi_j^{\prime}+2\phi_i^{\prime} \phi_j^{\prime}\phi_j^{\prime\prime}=V_{\phi_i}\,,
\ee
and the first-order and stressless equations are
\be
\phi_i^{\prime}\phi_j^{\prime}\phi_j^{\prime}=W_{\phi_i} \mbox{\;\;\;\;\;}\frac34\left(\phi_j^{\prime}\phi_j^{\prime}\right)^2=V.
\ee
Here we have
\be
\phi_i^{\prime}={W_{\phi_i} }\,\left(W_{\phi_j} W_{\phi_j}\right)^{-\frac13}\,,
\ee
and
\be
V=\frac34\left(W_{\phi_j}W_{\phi_j}\right)^{\frac23}\,.
\ee
Also, the stability equations become
\begin{equation}
-\left(a_{ij}\eta_i^{\prime}\right)^{\prime}-c_{ij}\eta_j=\omega^2 \phi_j^\prime\phi_j^\prime\eta_i
\end{equation}
where $a_{ij}\!=\!\left(\delta_{ij} \phi_k^\prime\phi_k^\prime+2 \phi_i^\prime\phi_j^\prime\right)$, $b_{ij}\!=\!0$ and $c_{ij}\!=\!-V_{\phi_i\phi_j}$.

We consider another model, defined by
\be
{\cal L}={\cal X}+\alpha{\cal X}\left|{\cal X}\right|-V(\phi_i)\,,
\ee
The equations of motion are
\be
\phi_i^{\prime\prime}\left(1+\alpha\phi_j^{\prime} \phi_j^{\prime}\right)+2\alpha\phi_i^{\prime}\phi_j^{\prime}\phi_j^{\prime\prime}=V_{\phi_i}\,,
\ee
and  the first-order and stressless equations are
\be
\phi_i^{\prime}\left(1+\alpha\phi_j^{\prime} \phi_j^{\prime}\right)=W_{\phi_i}\,,
\ee 
\be
\frac12 \phi_i^{\prime} \phi_i^{\prime}\left(1+\frac32\alpha \phi_j^{\prime} \phi_j^{\prime}\right)=V\,.
\ee
For $\alpha<<1$, we can get results up to first-order in $\alpha$; from the above equations we have
\be
\phi_i^{\prime}={W_{\phi_i} }\left(1-\alpha W_{\phi_j}W_{\phi_j}\right)\,,
\ee
and
\be
V=\frac12W_{\phi_j}W_{\phi_j}\left(1-\frac{\alpha}{2}W_{\phi_i}W_{\phi_i}\right)\,.
\ee
Also, the stability equations become
\begin{equation}
-\left(a_{ij}\eta_i^{\prime}\right)^{\prime}-c_{ij}\eta_j=\omega^2 \left(1+\alpha\phi_j^\prime\phi_j^\prime\right)\eta_i,
\end{equation}
with $a_{ij}\!=\!\delta_{ij}\left( 1\!+\!\alpha\phi_k^\prime\phi_k^\prime\right)\!+\!2\alpha \phi_i^\prime\phi_j^\prime$, $b_{ij}\!\!=\!\!0$ and $c_{ij}\!\!=\!\!-V_{\phi_i\phi_j}$.

We can use the recipe given previously to rewrite the stability equations above as Schroedinger-like equations, but this is out of the scope of the present work.

\section{Twinlike Models}

In this Section we focus on twinlike models. The main feature of twinlike models is that two distinct models may support the same solution, with the very same energy density. In the following, we present two distinct formalisms to construct twinlike models and examine the corresponding linear stability.

\subsection{Formalism I}

We consider that
\begin{equation}\label{eq15}
{\cal L}=-V(\phi_i)F(Y)\,,
\end{equation}
where $Y$ is defined as
\begin{equation}\label{eq16}
Y=-\frac{1}{2}\frac{\partial_\mu\phi_j\partial^\mu\phi_j}{V}\,.
\end{equation}
We note that for $F(Y)=1+Y$ we obtain the standard model, described by Eq.~\eqref{lsm}. The presence of $V$ in \eqref{eq15} and the numerator in \eqref{eq16} are important to avoid singuralies in the generalized  models, due to the zeroes of the potential.

The equation of motion is given by
\begin{equation}\label{eq17}
\partial_\mu\left(F_Y\partial^\mu\phi_i\right)+(F-YF_Y)V_{\phi_i}=0\,,
\end{equation}
and the energy-momentum tensor has the form
\begin{equation}
T_{\mu\nu}=g_{\mu\nu}V(\phi_i)F(Y)+F_Y\partial_\mu \phi_i\partial_\nu\phi_i\,,
\end{equation}
where $F_Y=dF/dY$.

As before, here we are interested in static field configurations; so, the equations of motion become
\begin{equation}\label{eq18}
-\left(F_Y\phi_i^{\prime}\right)^{\prime}+ (F-YF_Y)V_{\phi_i}=0\,.
\end{equation}

Moreover, for static solutions, the energy-momentum tensor gives
\bes
\begin{eqnarray}
T_{00}&=&VF\,,\label{eq19.1} \\
T_{11}&=&-V(F-2YF_Y)\,.\label{eq19.2}
\end{eqnarray}
\ees
The above Eq.~\eqref{eq18} can be integrated once to give
\begin{equation}\label{eq20}
2YF_Y-F=\frac{C}{V}\,.
\end{equation}
Again, $C$ is a constant. Furthermore, we have
\begin{equation}\label{eq21}
Y=\frac{1}{2}\frac{\phi_i^{\prime}\phi_i^{\prime}}{V(\phi_i)}\,.
\end{equation} 
and the Eq.~\eqref{eq20} can be written in the form
\begin{equation}\label{eq22}
\phi_i^{\prime}\phi_i^{\prime}=2G\left(\frac{C}{V}\right) V(\phi_i)\,,
\end{equation}
where $G=G(C/V)$ is an inversible function, with inverse $G^{-1}(Y)=2YF_Y-F$. 

For stressless solutions, that is, for $C=0$, we have that $2YF_Y\!=\!F$ and if we assume that $G(0)=c$, with $c$ a real constant, we find that $Y=c$. With this result, we can rewrite Eq.~$\eqref{eq22}$ in the form
\begin{equation}
\phi_i^{\prime}\phi_i^{\prime}=2c \,V({\phi_i})\,.
\end{equation}
If we consider that $V(\phi)=\frac12W_{\phi_i}W_{\phi_i}$, we get
\begin{equation}\label{eq23}
\phi_i^{\prime}=\sqrt{c} \;W_{\phi_i}\,.
\end{equation}
Here we note that the solution $\phi_i(x)$ of this equation is the same solution $\phi^{s}_{i}(x)$ of the Eq.~\eqref{eq4}, which appears for the standard model, with the position changed as $x \to \sqrt{c}\; x$. This means that we can write
\be
\phi_i(x) = \phi^s_{i}(\sqrt{c}\,x), \label{24}\,,
\ee
and now the thickness of the solution is given by
\be
\delta=\delta^s/\sqrt{c}\,.
\ee
{Thus, the solution is thicker or thinner, depending on the value of $c$ being lesser or greater than unit. We also note that $c$ cannot be negative;  and more, only stressless solutions have the specific form, given by Eq.~ \eqref{24}.}

The energy density of the stressless solution \eqref{eq23} gets to the form
\begin{equation}\label{eq25}
\rho(x)=\frac{F(c)}{2\sqrt{c}}\phi_i^{\prime}W_{\phi_i}=\frac{F(c)}{2\sqrt{c}}\frac{dW}{dx}\,.
\end{equation}
The energy is then
\ben
E&=&\frac{F(c)}{2\sqrt{c}}\int_{-\infty}^\infty dW\nonumber\\
&=&\frac{F(c)}{2\sqrt{c}}E_s\,,
\een
where $E_s$ is the energy giving by \eqref{eq6}. For $c=1$, we have to impose 
\bes\label{twinconditions}
\be \label{eq26}
F(1)=2\,,
\ee
in order to make the Eqs.~$\eqref{eq23}$ and $(\ref{eq25})$ identical to the Eqs.~$\eqref{eq4}$ and $(\ref{eq5})$, respectively. This also imposes that 
 \be \label{eq27}
 F_Y(1)=1\,.
 \ee\ees
The Eqs.~\eqref{eq26} and \eqref{eq27} are the general restrictions on $F(Y)$, to make the model defined by \eqref{eq15} twin of the standard model \eqref{lsm}. They are the conditions
to make the models twins, as pointed out in Ref.~\cite{tro}. There is another condiction, that makes the models to have the very same stability, which we discuss below. This was first introduced in the third paper in Ref.~\cite{PBSC}, and further explored in the fourth paper in Ref.~\cite{PBSC} and in other more recent investigations.

\subsubsection{Linear Stability}
	
Let us again investigate linear stability by introducing small fluctuations $\eta_i(x,t)$ in the static solution $\phi_i(x)$. As usual,  considering $\eta_i(x,t)=\eta_i(x)\cos(\omega t)$, from \eqref{stab1} and \eqref{eq15} with  \eqref{eq21}, we obtain 
\ben\label{Fnstab}
&&-\Big[F_Y\,\eta_i^{\prime}+2YF_{YY}\frac{\phi_i^{\prime}\phi_j^{\prime}}{\phi_k^{\prime}\phi_k^{\prime}}\,\eta_j^{\prime}\Big]^{\prime}\nonumber\\
&&-2{YF_{YY}}\frac{\left({\phi_i}^{\prime\prime}\,\phi_j^{\prime}-{\phi_j^{\prime\prime}}\,\phi_i^{\prime}\right)}{\phi_k^{\prime}\phi_k^{\prime}}\,\eta_j^{\prime}\nonumber\\
&&+\Bigg[(F-YF_Y)V_{\phi_i\phi_j}+2Y^2F_{YY}\frac{\phi_i^{\prime}\phi_l^{\prime}}{\phi_k^{\prime}\phi_k^{\prime}}V_{\phi_j\phi_l}+\nonumber\\
&&+4Y F_{YY}\Bigg(\frac{\phi_i^{\prime}\phi_j^{\prime}}{\phi_k^{\prime}\phi_k^{\prime}}-\frac{\phi_j^{\prime\prime}\phi_l^{\prime\prime}\phi_i^{\prime}\phi_l^{\prime}}{(\phi_k^{\prime}\phi_k^{\prime})^2}\Bigg)\Bigg]\,\eta_j \nonumber\\
&&=\omega^2 F_Y\,\eta_i\,.
\een

For the standard model $F=1+Y$, so we get
\be\label{nstab}
-\eta_i^{\prime\prime}+\left[V_{\phi_i\phi_j}\right]_{\phi_k=\phi_{sk}}\eta_j=\omega^2\,\eta _i\,,
\ee
as expected.\\

In the general situation $F=F(Y)$, using the stressless solutions of \eqref{eq20} we obtain, in the case of a single field,
\be
-\eta^{\prime\prime}+c\left[V_{\phi\phi}\right]_{\phi=\phi_s\left(\sqrt{c}\,x\right)}\eta=\frac{\omega^2}{A^2}\,\eta\,,
\ee
where
\be
A^2=\frac{F_Y+2YF_{YY}}{F_Y}\,.
\ee
In this case, if we have $A^2>0$, the two models have the same behavior under linear stability. See, e.g., the third paper in Ref.~\cite{PBSC}

In the more general case of several fields, in order to reduce the relation \eqref{Fnstab}  to equation \eqref{nstab}, we have to have the two conditions $F(1)=2$ and $F_Y(1)=1$, and another one, given by  $F_{YY}(1)=0$. These three condictions make the models twin, with the very same fluctuation spectra. For instance, one can write
\be\label{NN}
F_n(Y)= A_0+\sum^n_{k=1}\frac{A_k\,Y^k}{k}\,,\,\,\,\,\,n\geq 3\,,
\ee
where $A_k$ are real constants, and $A_0\neq 1$ for $n=3$. The models
defined by means of this $F$, satisfy the above conditions for 
\bes
\ben
A_1&=&4-3A_0-\frac12\sum^n_{k=4}\frac{(k-3)(k-2)A_k}{k}\,,\\
A_2&=&-6(1-A_0)+2\sum^n_{k=4}\frac{(k-3)(k-1)A_k}{k}\,,\\
A_3&=&3(1-A_0)+\frac32\sum^n_{k=4}\frac{(k-2)(k-1)A_k}{k}\,.
\een
\ees
One can have an infinity series, if the sum $\sum^{\infty}_{k=4}{A_k}/{k}$ converges, which is the case for
$A_k=1$, for $k\geq4$.

\subsubsection{Illustration}

We illustrate the general situation with $n=3$. Here we get
\be
F_3(Y)=A_0+A_1Y+\frac{A_2}{2}Y^2+\frac{A_3}{3}Y^3
\ee
We follow the general procedure to write $A_1$, $A_2$ and $A_3$ in terms of $A_0$, to obtain
\be
{F_3(Y)=A_0+(4-3A_0)Y-3(1-A_0)Y^2+(1-A_0)Y^3},{\;\;\;\;\;\;}
\ee
and
\ben\label{n3}
{\cal L}&=&-A_0V+(4-3A_0){\cal X}\nonumber\\
&&+3(1-A_0)\frac{{\cal X}^2}{V}+(1-A_0)\frac{{\cal X}^3}{V^2}.
\een
As informed below Eq.~\eqref{NN}, we cannot take $A_0=1$ for $n=3$. This would give $F=1+Y$ and ${\cal L}={\cal X}-V$, leading us back to the standard model.

We can also consider the case $n=4$. We have
\be
F_4(Y)=A_0+A_1Y+\frac{A_2}{2}Y^2+\frac{A_3}{3}Y^3+\frac{A_4}{4}Y^4.
\ee
It can be written as
\ben
F_4(Y)&=&A_0+A_1Y+3(3-2A_0-A_1)Y^2\nonumber\\
&&\!+\!(8A_0\!+\!3A_1\!-\!11)Y^3\!+\!(4\!-\!3A_0\!-\!A_1)Y^4,{\;\;\;\;\;}
\een
and now the Lagrange density becomes
\ben\label{n4}
{\cal L}&=&-A_0V+A_1{\cal X}-3(3-2A_0-A_1)\frac{{\cal X}^2}{V}\nonumber\\
&&+\!(8A_0\!+\!3A_1-11)\frac{{\cal X}^3}{V^2}\!-\!(4-3A_0-A_1)\frac{{\cal X}^4}{V^3}.\;\;\;\;\;\;\;
\een
If we choose $A_0=A_1=1$ we get back to the standard model. If we choose $A_1=4-3A_0$ we get back to the previous case, with $n=3$.
In the general case, however, both models \eqref{n3} and \eqref{n4} are twins of the standard model, for any valid potential; so, they are also twins of each other.
Thus, we have the case of triplets, and we can continue the process to find quads, quints, and in general multiple twinlike models.
We can consider the two-field model \cite{bnrt}
\be\label{ww}
W=\phi_1-\frac13\phi_1^3-r\phi_1\phi_2^2.
\ee
We take $r$ in the interval $r\in(0,1)$, and the potential
\be
V(\phi_1,\phi_2)=\frac12(1-\phi_1^2-r\phi_2^2)^2+2r^2\phi_1^2\phi_2^2,
\ee
gives rise to very nice defect solutions in the standard case \cite{bnrt}, which can be used to define generalized models like the previous ones, in \eqref{n3} and \eqref{n4}, with the very same defect solutions. We can consider other two-field models; see, e.g., Ref.~\cite{O1,O2}. We can also consider three-field models, as the ones used in \cite{3f}, for instance; this would lead us to other twinlike models.

Another example is obtained if one consider
\be
F(Y)= 2+\frac{a}{\alpha}\sin(\alpha Y)-\frac{b}{\alpha}\cos(\alpha Y)\,,
\ee
where $a,b$ and $\alpha$ are real constants. The models
defined by means of this $F$, satisfy the conditions $F(1)=2$, $F_Y(1)=1$  and $F_{YY}(1)=0$, for $\alpha={\rm arctan}(b/a)$ and $a^2+b^2=1$.
Particularly, for $b=0$ we have $\alpha=(2m+1)\pi/2$ ($m=0,\pm1,\pm2,...$); and for $a=0$ we obtain $\alpha=m\pi$ ($m=\pm1,\pm2,...$).

As we have just seen, all the models introduced in this subsection can be seen as twinlike models, and they may also have the same fluctuation spectra.

\subsection{Formalism II}

We will now develop a new formalism which allows to obtain twinlike models. For this,  we assume that the Lagrange density has the form
\begin{equation}\label{eq15ii}
{\cal L}=-\sum_j\frac12 W_{\phi_j}^2F^j\,,
\end{equation}
where $F^j$ depends on $Y_j$, and again $j=1,2,\ldots,N$. However, differently from the previous formalism, we now assume that there exist $N$ distinct functions $F^j$ and $Y_j$, where each $Y_j$ is defined as
\begin{equation}\label{17aii}
Y_1=-\frac{\partial_\mu\phi_1\partial^\mu\phi_1}{W^2_{\phi_1}}\,,\,\,Y_2=-\frac{\partial_\mu\phi_2\partial^\mu\phi_2}{W^2_{\phi_2}}\,,\,\,...
\end{equation}
We note that for $F^j=1+Y_j$ we obtain the standard model introduced by \eqref{lsm} and \eqref{vsm}.

Here, the equations of motion are given by, for $i=1,2,...,N$,
\begin{equation}
\partial_\mu\left(F^i_{Y_i}\partial^\mu\phi_i\right)+\sum_j(F^j-Y_jF^j_{Y_j})W_{\phi_j}W_{\phi_j\phi_i}=0,
\end{equation}
where $F^j_{Y_j}=dF^j/dY_j$. Also, the energy-momentum tensor is
\begin{equation}
T_{\mu\nu}=\sum_j\left(F^j_{Y_j}\partial_\mu \phi_j\partial_\nu\phi_j+\frac12g_{\mu\nu}W^2_{\phi_j}F^j\right)\,,
\end{equation}

For static field configurations, the equations of motion become
\begin{equation}\label{eq18ii}
\left(F^i_{Y_i}\phi_i^{\prime}\right)^{\prime}-\sum_j(F^j-Y_jF^j_{Y_j})W_{\phi_j}W_{\phi_j\phi_i}=0\,,
\end{equation}
and the energy-momentum tensor gives
\bes
\begin{eqnarray}
T_{00}&=&\frac12 \sum_j\,W^2_{\phi_j} F^j\label{t00ii}\,, \\
T_{11}&=&-\frac12\sum_j\,  W^2_{\phi_j} (F^j-2Y_jF^j_{Y_j})\,.\label{t11ii}
\end{eqnarray}
\ees

For every $i$, the set of $N$ equations \eqref{eq18ii} can be integrated once to give
\begin{equation}\label{eq20ii}
2Y_iF^i_{Y_i}-F^i=\frac{2C_i}{W^2_{\phi_i}}\,,
\end{equation}
where the several $C_i$ represent real constants. From Eq.~\eqref{17aii} we have
\be\label{eq21ii}
Y_1=\frac{\phi_1^{\prime\,2}}{W^2_{\phi_1}}\,,\,\,Y_2=\frac{\phi_2^{\prime\,2}}{W^2_{\phi_2}}\,,\,\,...
\end{equation} 
and the set of $N$ equations, Eq.~\eqref{eq20ii}, can be written in the form
\begin{equation}\label{eq22ii}
\phi_i^{\prime\,2}=G_i\left(\frac{2C_i}{W^2_{\phi_i}}\right) W^2_{\phi_i}\,,
\end{equation}
for each $i$, where $G_i$ is a function with inverse $G_i^{-1}(Y_i)=2Y_iF^i_{Y_i}-F^i$. 

For stressless solutions, that is, for $C_i=0$, we have that $2Y_iF^i_{Y_i}=F^i$ and if we assume that $G_i(0)=c_i$, with $c_i$ representing real constants, we find that $Y_i=c_i$.  From Eq.~$\eqref{eq22ii}$, with this result, we get
\begin{equation}\label{eq23ii}
\phi_i^{\prime}=\sqrt{c_i} \;W_{\phi_i}\,.
\end{equation}
Here, we note that the solution $\phi_i(x)$ of this equation is the same solution $\phi^s_{i}(x)$ of the Eq.~\eqref{eq4}, which appears for the standard model, with the position changed as $x \to \sqrt{c_i}\; x$. This means that we can write
\be
\phi_i(x) = \phi^s_{i}(\sqrt{c_i}\,x), \label{24ii}\,
\ee
and now the thickness of the several fields obey 
\be
\delta_i=\delta^s_{i}/\sqrt{c_i}\,.
\ee
It is thicker or thinner, depending on the value of $c_i$ being lesser or greater than unit. We also note that $c_i$ cannot be negative;  and more, only stressless solutions have the specific form, given by Eq.~ \eqref{24ii}.

The energy density of the stressless solution \eqref{eq23ii} gets to the form
\begin{equation}\label{eq25ii}
\rho(x)=\sum_i\frac{F^i(c_i)}{2\sqrt{c_i}}\phi_i^{\prime}W_{\phi_i}=\sum_i\frac{F^i(c_i)}{2\sqrt{c_i}}\frac{dW}{dx}\,.
\end{equation}
Then, the energy is given by
\be
E=\sum_i\frac{F^i(c_i)}{2\sqrt{c_i}}\int_{-\infty}^\infty dW=\sum_i\frac{F^i(c_i)}{2\sqrt{c_i}}E_s\,,
\ee
where $E_s$ is the energy giving by \eqref{eq6}. For $c_i=1$, we have to impose 
\bes\label{twinconditions}
\be \label{eq26ii}
F^i(1)=2\,,
\ee
in order to identify the Eqs.~$\eqref{eq23ii}$ and $(\ref{eq25ii})$ to the Eqs.~$\eqref{eq4}$ and $(\ref{eq5})$, respectively. This also imposes that 
 \be \label{eq27ii}
 F^i_{Y_i}(1)=1\,.
 \ee
\ees
For each $i$, the Eqs.~\eqref{eq26ii} and \eqref{eq27ii} are the general restrictions on $F^i$, to make the generalized model twin of the standard model \eqref{lsm}.
These are the two conditions, to make the models twins of each other.

\subsubsection{Linear Stability}

Let us again investigate linear stability by introducing small fluctuations $\eta_i(x,t)$ in the static solution $\phi_i(x)$. As before, we consider $\eta_i(x,t)=\eta_i(x)\cos(\omega t)$, and from \eqref{stab1} and \eqref{eq15ii} with  \eqref{eq21ii}, for stressless condition, we obtain 
\ben\label{Fnstabii}
&&-\Big[\Big(F^i_{Y_i}+2Y_iF^i_{Y_iY_i}\Big)\eta_i^{\prime}\Big]^{\prime}\nonumber\\
&&+2\sum_j\Big(Y_i^{3/2}F^i_{Y_iY_i}-Y_j^{3/2}F^j_{Y_jY_j}\Big)\,W_{\phi_i\phi_j}\,\eta_j^{\prime}\nonumber\\
&&+\sum_{j,l}\Bigg[(F^l-Y_l F^l_{Y_l}+2Y_l^2F^l_{Y_lY_l})W_{\phi_i\phi_l}W_{\phi_j\phi_l}\nonumber\\
&&+(F^l-Y_lF^l_{Y_l}+2Y_i^{3/2}Y_l^{1/2}F^i_{Y_iY_i})W_{\phi_l}W_{\phi_i\phi_j\phi_l}\Bigg]\,\eta_j \nonumber\\
&&=\omega^2 F^i_{Y_i}\,\eta_i\,.
\een
This is a general result. We note that for the standard model we have to use $F^i=1+Y_i$; in this case we get
\be\label{nstabii}
-\eta_i^{\prime\prime}+\left[V_{\phi_i\phi_j}\right]_{\phi_j=\phi^s_j}\eta_j=\omega^2\,\eta _i\,,
\ee
where $V=(1/2)\sum_j W^2_{\phi_j}$, as expected.

In the current case, we have an interesting result to highlight. It refers to the two distinct ways to make the generalized model to behave as the standard model, concerning linear stability. The first possibility refers to reducing the Eq.~\eqref{Fnstabii} to Eq.~\eqref{nstabii} by imposing the additional condition $F^i_{Y_iY_i}(1)=0$, like in the previous case. The other possibility appears for $Y_1=Y_2=...=Y_N=Y=c$, when we take the same functional form for the functions $F^j$; that is, we take $F^1=F^2=...=F^N=F$, for the several fields. In this case, the several Eq.~\eqref{Fnstabii} reduce to
\be
-\eta_i^{\prime\prime}+c\sum_j \left[V_{\phi_i\phi_j}\right]_{\phi_j=\phi^s_j\left(\sqrt{c}\,x\right)}\eta_i=\frac{\omega^2}{A^2}\,\eta_i\,,
\ee
where
\be
A^{2}=\frac{F_{Y}+2YF_{YY}}{F_{Y}}\,,
\ee
Thus, for $A^2>0$, the two models have the same stability behavior, as it happens in the case of one field. This is the strong twin condition, that makes the models to have the very same stability. See, e.g., the third and fourth papers in  Ref.~\cite{PBSC}.

\subsubsection{Illustration}

Here we consider a two-field model, with
\be
{\cal L}=-\frac12W_{\phi}^2F^1(Y_1)-\frac12W_{\chi}^2F^2(Y_2).
\ee
We take as $F^1$ and $F^2$ the previous $F_3$ and $F_4$, that is, we consider
\ben
F^1(Y_1)&=&B_0+(4-3B_0)Y_1\nonumber\\
&&-3(1-B_0)Y_1^2+(1-B_0)Y_1^3\\
F^2(Y_2)&=&A_0+A_1Y_2+3(3-2A_0-A_1)Y_2^2\nonumber\\
&&+(8A_0+3A_1-11)Y_2^3\nonumber\\
&&+(4-3A_0-A_1)Y_2^4,
\een
where $B_0$ and $A_0, A_1$ are real parameters. Also,
\be
Y_1=-\frac{2X_{11}}{W_{\phi}^2};  \mbox{\;\;\; \;\;\;}  Y_2=-\frac{2X_{22}}{W_{\chi}^2};
\ee
and
\be
X_{11}=\frac12\partial_\mu\phi\partial^\mu\phi;  \mbox{\;\;\; \;\;\;}  X_{22}=\frac12\partial_\mu\chi\partial^\mu\chi.
\ee
If we use $W$ as given by \eqref{ww}, with $\phi_1=\phi$ and $\phi_2=\chi$, we have another example of generalized model
of the class studied above. This two-field model is an explicit construction of twinlike models, and it may also have the very same fluctuation spectra.

\section{Conclusions}\label{end} 

In this work we studied generalized models, searching for kinklike structures under the first-order formalism, that is, for solutions that obey first-order differential equations that solve the equations of motion. This formalism was them used to investigate twinlike models, which are distinct models having the very same kinklike structure, with the same energy density
and the same linear stability. The main focus of the investigation was on the formal steps needed to write the general results.

We have introduced two distinct routes to get to generalized models. The first case considered the generalized model in the form
$$
{\cal L}=-V(\phi)F(Y)\,,
$$
with $Y$ defined as
$$
Y=-\frac{1}{2}\sum_j\frac{\partial_\mu\phi_j\partial^\mu\phi_j}{V}\,.
$$
The second case dealt with
$$
{\cal L}=-\sum_j\frac12 W_{\phi_j}^2F^j\,,
$$
where each $F^j$ depends on $Y_j$ alone, given by
$$
Y_j=-\frac{\partial_\mu\phi_j\partial^\mu\phi_j}{W^2_{\phi_j}}.
$$
The two routes are different, and allow for the construction of a diversity of models.

The twinlike models introduced in this work give rise to interesting defect structures, which are basically controlled by the potential and other functions, that depend on the derivative of the several scalar fields that specify each one of the models. There is a multiplicity of models of the twinlike type, each one of them having specific features, but  allowing for the same defect structure, with the same energy density and the very same linear stability.

A general feature of the generalized models is that they obey first-order differential equations, so a natural question to ask concerns the inclusion of fermions, to study if one
can find supersymmetric extensions of the above models \cite{Y}, to investigate the behavior of fermions under such generalized scenarios. This issue will be considered elsewhere.
Another line of investigation concerns cosmology, with the results of this work being of direct interest to describe multifield inflation and multifield defect networks, as suggested in Refs.~\cite{X,Z}. The case of multifield defect network is presently under investigation, following the lines of Ref.~\cite{X}. We intend to report on the issue in another work.

The authors would like to thank CAPES and CNPq for partial financial support.


\end{document}